\documentstyle[aps]{revtex}
\begin{document}

\draft

\title{The Choptuik spacetime as an eigenvalue problem}

\author{Carsten Gundlach}

\address{Physics Department, University of Utah, Salt Lake City,
UT84112, USA}
\address{LAEFF-INTA, PO Box 50727, 28080 Madrid, Spain
\footnote{Current address. Email: {\tt gundlach@laeff.esa.es}}}

\date{30 September 1994, revised 27 July 1995 and 8 September 1995}

\maketitle

\begin{abstract}
By fine-tuning generic Cauchy data, critical phenomena have recently
been discovered in the black hole/no black hole ``phase transition" of
various gravitating systems. For the spherisymmetric scalar field
system, we find the ``critical" spacetime separating the two phases by
demanding discrete scale-invariance, analyticity, and an additional
reflection-type symmetry.  The resulting nonlinear hyperbolic boundary
value problem, with the rescaling factor $\Delta$ as the eigenvalue,
is solved numerically by relaxation. We find $\Delta = 3.4439 \pm 0.0004$.
\end{abstract}


Recently, Choptuik \cite{Chop} has studied the gravitational collapse
of a real scalar field (massless or massive, minimally or conformally
coupled) in spherical symmetry, using an adaptive mesh
refinement numerical technique which allows him to study details on
very small spacetime scales.  To describe his results concisely, we
invoke coordinates $\{p,\bar p\}$ on the phase space of the
spherisymmetric gravitating scalar field, where $p$ is any smooth
coordinate such that $p=0$ is the hypersurface which divides
black-hole from no-black-hole spacetimes, while $\bar p$ denotes the
remaining coordinates.  Choptuik's results strongly indicate the
following conjectures:

(1) For any choice of coordinate system $\{p,\bar p\}$, the mass of
sufficiently small black holes is given by $M= f(\bar p)\, p^\gamma$
(``scaling''), where $\gamma\sim 0.37$ is a universal exponent.

(2) There is a ``critical solution" $\{p=0, \bar p=\bar p_*(t)\}$,
which acts as an intermediate attractor in a thin sheet surrounding
the $p=0$ hypersurface on both sides (``universality").

(3) This solution shows a discrete homotheticity (``echoing''), or
scale invariance, to be defined more precisely below, with a
logarithmic rescaling factor $\Delta \sim 3.44$.

More recent research indicates that these properties hold for other
self-gravitating systems. Universality, echoing with ($\Delta\sim
0.6$) and scaling (with $\gamma \sim 0.37$) were found in collapse of
axisymmetric gravitational waves \cite{AbrEv}.  Universality,
continuous self-similarity and scaling (with $\gamma \sim 0.36$) were
found in perfect fluid collapse with $p=\rho/3$ \cite{EvCol}. The
exactly self-similar solution was calculated as an eigenvalue problem
\cite{EvCol}, and the critical exponent calculated to high precision
($\gamma=0.3558019$) by perturbing it \cite{Koike}. Critical exponents
for other values of $k$ in $p=k\rho$ were calculated in this manner
\cite{Maison} and confirmed in collapse calculations \cite{Evans},
with values strongly dependent on $k$.  Self-similar solutions were
also calculated for a complex scalar field \cite{HE1} and an axion-dilaton
combination \cite{EHH}. A value ($\gamma=0.387106$) of the critical
exponent for the complex scalar field was derived by perturbing the
self-similar solution \cite{HE2}, but the latter is apparently not an
attractor \cite{pc}.

Universality, scale invariance and critical exponents indicate an
exciting new connection between renormalisation group theory and
classical general relativity. $\gamma$ appears to vary from one
physical system to another, but its values for vacuum or trace-free
matter are remarkably similar in the examples found so far.

In this Letter, we impose echoing, and an additional reflection-type
symmetry, in our ansatz, together with analyticity, and solve the
resulting nonlinear hyperbolic eigenproblem, instead of evolving and
fine-tuning Cauchy data. In the language of renormalisation group
theory, we find a fixed point of gravitational collapse under a
rescaling of space and time by solving the renormalisation group
equations. In a future paper we intend to calculate $\gamma$ by
perturbing around the fixed point, along the lines of
\cite{Koike,Maison,HE2}.

The Einstein equations we consider here are
\begin{equation}
G_{ab}=8\pi G\, \left(\phi_{,a}\phi_{,b}-{1\over2}g_{ab}
\phi_{,c}\phi^{,c}\right),
\end{equation}
in spherical symmetry.  The matter equation $\phi_{,c}^{\ \ ;c}=0$
follows as a Bianchi identity.
Following Choptuik, we define the metric as
\begin{equation}
\label{metric}
ds^2=-\alpha(r,t)^2\,dt^2+a(r,t)^2\,dr^2+r^2\,
(d\theta^2+\sin^2\theta\,d\varphi^2),
\end{equation}
where the remaining gauge freedom is fixed by the condition
$\alpha(r=0,t)\equiv 1$, and auxiliary matter fields as
\begin{equation}
 X(r,t)=\sqrt{2\pi G}\ {r\over a}\phi_{,r},\qquad
Y(r,t)=\sqrt{2\pi G}\ {r\over \alpha}\phi_{,t}.
\end{equation}

The symmetry of the attractor observed by Choptuik can be expressed in
coordinate language as $Z(r,t)=Z(re^\Delta,te^\Delta)$, where $Z$
stands for any one of $\alpha$, $a$, $X$ and $Y$, and $\Delta \sim
3.44$ is a constant.  Here the zero of $t$ has been adjusted so that
$(0,0)$ is the accumulation point of the echos. We introduce auxiliary
(nonmetric) coordinates where the symmetry appears as a simple
periodicity of the $Z$:
\begin{equation}
\label{echoing}
\tau\equiv\ln(\pm t),
\quad\xi\equiv\ln(\pm r/t),
\quad Z(\xi,\tau)=Z(\xi,\tau+\Delta).
\end{equation}
In the following, we use the upper sign ($t>0$). The equations for the
lower sign are obtained by changing the sign of $Y$ in all following
equations.

Geometrically, the symmetry can be described as a discrete
homotheticity: when we Lie-drag $g_{ab}$ along the vector field
$\partial/\partial \tau$ by the distance $\Delta$, we obtain
$g_{ab}e^{2\Delta}$, while $T_{ab}$ is mapped to $T_{ab}$. The vector
field along which we Lie-drag is not unique, however, because we only
ever consider the effect of Lie-dragging the finite distance
$\Delta$. We parameterize this arbitrariness by introducing a free
periodic function $\xi_0(\tau)$ into the coordinate system such that
the vector involved in the symmetry is still $\partial/\partial
\tau$. At the same time, for clarity of presentation, we absorb
$\Delta$ into the coordinate $\tau$. We therefore define the
coordinates in which we are going to work as
\begin{equation}
\varphi\equiv 2\pi\tau/\Delta,\quad \zeta\equiv\xi-\xi_0(\tau),
\quad Z(\zeta,\varphi+2\pi)=Z(\zeta,\varphi).
\end{equation}

Evans and Coleman \cite{EvCol} found the critical spacetime of
spherical fluid collapse by imposing continuous homotheticity. In our
coordinates, $\alpha$, $a$, and the fluid variables corresponding to
$X$ and $Y$ are then functions of $\zeta$ alone, the Einstein and
matter equations are reduced to a system of nonlinear ODEs, and the
solution is uniquely specified by regularity conditions
\cite{EvCol,Selfsim}. Here we use a similar approach in order to find
Choptuik's critical spacetime of scalar field collapse.  (After this
work was begun, the same approach was used to calculate (continuously)
self-similar solutions for the complex scalar field \cite{HE1} and an
axion-dilaton combination \cite{EHH}.)

We solve the field equations for the $\zeta$-derivatives of the fields
$Z$ as functions of these fields and their $\varphi$-derivatives. It is
convenient to use the new field $g\equiv e^{\xi_0(\varphi)} a/\alpha$
instead of $\alpha$, and $X_\pm\equiv X\pm Y$ instead of $X$ and
$Y$. The resulting equations are
\begin{eqnarray}
\label{adot}
a_{,\zeta}=&&{1\over 2}a\left[
(1-a^2)+a^2\left(X_+^2+X_-^2\right)\right],\\
\label{gdot}
g_{,\zeta}=&&g(1-a^2),\\
\label{xpdot}
X_{+,\zeta}=&&{B_+ \over 1+D},\\
\label{xmdot}
X_{-,\zeta}=&&{B_- \over 1-D},
\end{eqnarray}
where we have introduced the abbreviations
\begin{eqnarray}
z&&\equiv \left(1+ {2\pi\over\Delta}{d\xi_0\over
d\varphi}\right)^{-1},
\quad
D\equiv z^{-1}e^\zeta g,\\
B_{\pm}&&\equiv{1\over2}(1-a^2)X_\pm -a^2 X_\mp^2 X_\pm -X_\mp
\pm z{2\pi\over\Delta}DX_{\pm,\varphi}.
\end{eqnarray}
There is also one equation containing only the $Z$ and $Z_{,\varphi}$,
\begin{equation}
\label{constraint}
z{2\pi\over\Delta}{a_{,\varphi}\over a}={1\over2}\left[(1-a^2)
+a^2\left(X_+^2+X_-^2\right)+a^2D^{-1}\left(X_+^2-X_-^2\right)\right].
\end{equation}
It acts as a constraint, which is conserved
by the four ``evolution equations" above.

For small enough $\zeta$ these equations define a constrained Cauchy
problem, with $\zeta$ playing the role of time, on the cylinder
obtained by identifying $\varphi$ with period $2\pi$. At
$\zeta=-\infty$, corresponding to $r=0$, we set the boundary
conditions $a=1$ (regularity of the metric) and $\alpha=1$ (coordinate
condition). Expanding the field equations in powers of $e^\zeta$, we
find that data obeying these conditions are determined by
$\xi_0(\varphi)$ and one more free function $Y_0(\varphi)$, which is
defined by the expansion
\begin{equation}
Y(\varphi,\zeta)\equiv Y_0(\varphi)e^{\xi_0(\varphi)}\, e^{\zeta}
+O\left(e^{3\zeta}\right).
\end{equation}

As $\zeta$ increases, the Cauchy problem eventually becomes
degenerate, when $D=1$. In analogy to the ``sonic point" of the
ODEs describing continuously homothetic spacetimes
\cite{EvCol,Selfsim}, we call this line the ``sonic line". The
equation of radial null geodesics is
\begin{equation}
{d\zeta\over d\varphi}=-{\Delta\over 2\pi z}\left(-1\pm D^{-1}\right).
\end{equation}
The sonic line is therefore the set of points where a null geodesic
touches a surface of constant $\zeta$. (In general, it would be a
matter characteristic, but for our choice of matter these are
identical with the null geodesics.)  The solution can be uniquely
continued across the sonic line when we impose analyticity. As a
technical simplification, we make use of the coordinate freedom in
$\xi_0$ by moving the sonic line to $\zeta=0$.  Then we can enforce
analyticity simply by expanding in powers of $\zeta$. We find that
regular data near $\zeta=0$ can be expressed in terms of
$\xi_0(\varphi)$ and one more free function $X_{+0}(\varphi)$, which
is defined as
\begin{equation}
X_{+0}(\varphi)\equiv X(\zeta=0,\varphi)+Y(\zeta=0,\varphi).
\end{equation}
We have now formulated a hyperbolic boundary value problem on a
rectangle with two sides identified (a finite cylinder) in $1+1$
dimensions.  We have three independent fields, for example $X_+$,
$X_-$ and $g$. On the other hand there are three free functions in the
boundary data, minus one degree of freedom corresponding to
translations in $\varphi$, plus $\Delta$ as the eigenvalue of the
problem. By this count we expect solutions to be locally unique, with
a discrete spectrum.

We cut the number of degrees of freedom in half by imposing the
additional symmetry $Z(\varphi+\pi)=\pm Z(\varphi)$, with the $+$ sign
holding for $a$, $g$ and $\xi_0$, and the $-$ sign for $X_+$ and
$X_-$. It is consistent with the field equations and Choptuik's data
\cite{Chopdata}. Moreover, Choptuik observed that the massive scalar
field has the same attractor as the massless one considered here. The
necessary and sufficient condition for this is that $\phi$ remains
bounded, because the mass term in its stress tensor is then dominated
by the gradient squared term as $\phi$ varies on ever smaller
spacetime scales. For $\phi$ to remain bounded its derivatives $X$ and
$Y$ must have vanishing zero frequency Fourier components in
$\varphi$, which in turn requires that all their even frequencies
vanish, or else these could be combined to give a zero frequency
contribution in the evolution equations.  It follows in turn that $a$,
$g$ and $\xi_0$ must not have odd frequency components.

As we are dealing with smooth periodic functions, it is useful to
decompose all fields into their Fourier components with respect to
$\varphi$. Integration and differentiation are done in Fourier
components. Algebraic operations are done in $\varphi$ space, which
makes our algorithm pseudo-spectral. Due to the nonlinearity of the
problem, dealiasing turns out to be essential for stability. We
dealiase convolution sums by using a number of collocation points in
$\varphi$ equal to twice the number of Fourier components.

Because the number of variables is large and the problem is nonlinear
in an essential way (there are no regular solutions to the linearized,
no gravity, problem), any algorithm is likely to have only a small
region of convergence around any solution. We have therefore started
with an initial guess sufficiently close to Choptuik's critical
spacetime, in order to establish that this solution exists having the
echoing symmetry as an an exact symmetry, and that it is locally
unique, and to calculate it with higher precision than has been
possible by fine-tuning Cauchy data. A global search for all solutions
that may exist is desirable, but not possible with the present
algorithm.

Our manual input into the algorithm is limited to the following guess
for $Y_0$ and $\Delta$: $Y_0=-2.3\sin\varphi-0.6\sin 3\varphi$, and
$\Delta=3.44$. Here and in the following we fix the translation
invariance by defining $Y_0$ to have no $\cos\varphi$ component.  The
numbers were estimated from very near-critical collapse data made
available by Choptuik \cite{Chopdata}, after transformation to
coordinates $(\xi,\tau)$. In a first step, we begin with the very
rough guess $\xi_0=0$, and shoot from $\zeta=-\infty$ towards
increasing $\zeta$. When $D(\varphi,\zeta)$ first gets close to $1$ in
two points $\varphi$, and $X_{-,\zeta}$ is therefore about to become
singular in those points, we stop the evolution and calculate a new
value of $\xi_0$ that is designed to ``flatten'' $D(\varphi)$, i.e. to
make it roughly $D(\varphi)\sim 1$ for all $\varphi$ at that
$\zeta$. Then we shoot again, thus iteratively improving
$\xi_0$. After convergence, we read off $X_+$ at the endpoint of our
one-sided shooting, which by now is close to $\zeta=0$, and thus have
an initial guess for $X_{+0}$ as well.

As an intermediate step, we calculate an initial guess for the values of all
fields on a grid in $\zeta$ by shooting from both $\zeta=-\infty$ and
$\zeta=0$ to a fitting point, typically $\zeta=-1$. This involves a
Taylor expansion around the regular singular point $0$ as well
as around $-\infty$. Using this expansion and shooting from $\zeta=0$
transfers the bulk of the error in our improving solution away from
the point $\zeta=0$ to the fitting point $\zeta=-1$, making it easier
for the following step to handle.

In the last step, we go over to a standard relaxation algorithm
\cite{Press}. For the purpose of relaxation the independent variables
at each grid point in $\zeta$ are the odd Fourier components of $X_+$
and $X_-$ and the even components of $g$ and $\xi_0$. $a$ is not
considered as independent, but reconstructed at each step from the
other fields by solving equation (\ref{constraint}). Solution of this
ODE is by iteration of the corresponding integral equation. (The
constant component of $a$ has to be calculated separately.) Between
generic grid points in $\zeta$ we enforce the discretized
$\zeta$-derivatives
\begin{equation}
Z_{n+1}-Z_n=h\,F\left({Z_{n+1}+Z_n\over 2}\right).
\end{equation}
(The $\zeta$-derivative of $\xi_0$ is zero by definition). At the
boundary $\zeta=-\infty$ we enforce relations between $g$, $X$, and
$Y$ derived from expanding the field equations, and at $\zeta=0$ we
enforce $D=1$ and $B_-=0$.  The relaxation part of the algorithm is
much simpler than the shooting parts, but the latter appear to have a
larger region of convergence, thus serving as a stepping stone.

The boundary data of the solution have been tabulated in Table 1.  In
particular, the echoing period is $\Delta = ( 3.4439 \pm 0.0004)
$. The error bars have been obtained combining the results of three
different convergence tests. (1) We compare the results obtained for
different numbers $M$ of grid points in $\zeta$. As expected the
convergence is quadratic, over a wide range of $M$, but only up to
some maximal value. (2) The convergence of the tabulated numbers with
increasing number $N$ of Fourier components used in the calculation is
rapid (``spectral convergence'') for $N\ge 32$. (3) $\zeta=-\infty$ is
represented by a finite value of $\zeta$, using a Taylor expansion to
one beyond leading order in $\exp\zeta$. As expected, this convergence
is quadratic in $\exp\zeta$, over some range of $\zeta$. As long as
the difference between runs of different precision has the expected
functional form, we can use it to estimate the numerical error. The
tabulated data are from a run with $M=201$ equally spaced points in
the interval $-5\le\zeta\le 0$ and $N=64$ components (half of which
vanish) per function, compared with $-6\le\zeta\le 0$, $M=401$ and
$N=128$ respectively. The three sources of numerical error are
comparable for this choice.

We have compared the fields $a$, $\alpha$, $X$ and $Y$ with Choptuik's
data, after interpolating to the largest rectangular grid in $\tau$
and $\xi$ contained in both data sets, with $-3.2<\xi<1.3$. We have
evaluated the root mean square of the absolute difference point by
point of the fields $a$, $X$ and $Y$ (which are bounded and of order
one in the solution) and the relative difference in $\alpha$ (which is
unbounded above, but bounded below by $1$). After adjusting a
non-universal offset in $\tau$ between the data sets, this difference
is $3.9~10^{-2}$ for $\alpha$, and somewhat smaller for the other
fields. By comparison, the estimated root mean square pointwise error
in our data is $1.6~10^{-3}$ in $\alpha$, and $1.0~10^{-4}$ or less
for the other fields.  We have therefore improved the precision with
which the Choptuik spacetime is known by one to two orders of
magnitude, while $\Delta$ is now known to one part in $10^4$. Future
improvements are possible.

Data files of the solution are
available at {\tt http://www.laeff.esa.es/$\sim$gundlach/}.

The author would like to thank Silvano Bonazzola, Pat Brady, Matt Choptuik,
Chris Clarke, Charles Evans, Juan P\'erez Mercader, Richard Price and Jorge
Pullin for helpful and enjoyable conversations, and the Newton Institute,
Aspen Physics Center and DAMTP for hospitality while this work was
begun. This work was supported by NSF grant PHY9207225 and research funds
of the University of Utah, and subsequently by the Spanish Ministry of
Education and Science.

\vfill \break

\begin{table}
\caption{Decomposition of the boundary data in sines and cosines.
Their period is $\Delta = (      3.4439  \pm      0.0004) $.}
\begin{tabular}{rrrrr}
Component & $\xi_0$ & Component & $Y_0$ & $X_{+0}$ \\
\tableline
constant
& 	$(      1.5813  \pm      0.0007) \, $
&  $\cos\varphi$
&  $0$ \tablenote{by definition, to fix translation degree of freedom}
&  $(     -4.3831  \pm      0.0006) \, 10^{ -1}$
\\
$\cos 2\varphi$
& 	$(       6.658  \pm       0.006) \, 10^{ -2}$
&  $\sin\varphi$
&   $(      -2.364  \pm       0.006) \, $
&  $(     -3.2287  \pm      0.0008) \, 10^{ -1}$
\\
$\sin 2\varphi$
&$(      -1.577  \pm       0.002) \, 10^{ -1}$
&  $\cos 3\varphi$
&$(       -1.46  \pm        0.05) \, 10^{ -1}$
&$(        6.74  \pm        0.05) \, 10^{ -3}$
\\
$\cos 4\varphi$
&$(      -2.014  \pm       0.004) \, 10^{ -2}$
&  $\sin 3\varphi$
&$(       -9.52  \pm        0.08) \, 10^{ -1}$
&$(       1.017  \pm       0.001) \, 10^{ -1}$
\\
$\sin 4\varphi$
&  $(        -3.3  \pm         0.2) \, 10^{ -4}$
&  $\cos 5\varphi$
&  	$(       -1.12  \pm        0.05) \, 10^{ -1}$
&  $(       2.431  \pm       0.003) \, 10^{ -2}$
\\
$\cos 6\varphi$
&  	$(       1.979  \pm       0.005) \, 10^{ -3}$
&  $\sin 5\varphi$
&  	$(       -4.06  \pm        0.06) \, 10^{ -1}$
&  $(      -1.807  \pm       0.008) \, 10^{ -2}$
\\
$\sin 6\varphi$
&  	$(       2.249  \pm       0.008) \, 10^{ -3}$
&  $\cos 7\varphi$
&  	$(        -7.0  \pm         0.4) \, 10^{ -2}$
&  $(       -9.85  \pm        0.02) \, 10^{ -3}$
\\
$\cos 8\varphi$
&  	$(        1.37  \pm        0.01) \, 10^{ -4}$
&  $\sin 7\varphi$
&  $(       -1.73  \pm        0.04) \, 10^{ -1}$
&  $(       -2.14  \pm        0.02) \, 10^{ -3}$
\\
$\sin 8\varphi$
&  	$(      -8.186  \pm       0.004) \, 10^{ -4}$
&  $\cos 9\varphi$
&  	$(        -3.9  \pm         0.3) \, 10^{ -2}$
&  $(        1.76  \pm        0.01) \, 10^{ -3}$
\\
$\cos 10\varphi$
&  	$(      -1.886  \pm       0.002) \, 10^{ -4}$
&  $\sin 9\varphi$
&  $(        -7.3  \pm         0.2) \, 10^{ -2}$
& $(       3.116  \pm       0.005) \, 10^{ -3}$
\\
$\sin 10\varphi$
&  	$(         3.6  \pm         0.1) \, 10^{ -5}$
&  $\cos 11\varphi$
&  	$(        -2.1  \pm         0.2) \, 10^{ -2}$
& $(        3.97  \pm        0.02) \, 10^{ -4}$
\\
$\cos 12\varphi$
&  	$(        3.08  \pm        0.03) \, 10^{ -5}$
&  $\sin 11\varphi$
&  $(        -3.0  \pm         0.1) \, 10^{ -2}$
&   $(      -1.195  \pm       0.005) \, 10^{ -3}$
\\
$\sin 12\varphi$
&  	$(        1.73  \pm        0.02) \, 10^{ -5}$
&  $\cos 13\varphi$
&  	$(        -1.1  \pm         0.1) \, 10^{ -2}$
&$(       -4.05  \pm        0.01) \, 10^{ -4}$
\\
$\cos 14\varphi$
&  	$(       -4.05  \pm        0.09) \, 10^{ -6}$
&  $\sin 13\varphi$
&  $(       -1.24  \pm        0.06) \, 10^{ -2}$
& $(        1.93  \pm        0.02) \, 10^{ -4}$
\\
$\sin 14\varphi$
&  	$(      -1.371  \pm       0.002) \, 10^{ -5}$
&  $\cos 15\varphi$
&  	$(        -5.2  \pm         0.9) \, 10^{ -3}$
&  $(        1.53  \pm        0.01) \, 10^{ -4}$
\\
$\cos 16\varphi$
&  	$(       -3.05  \pm        0.03) \, 10^{ -6}$
&  $\sin 15\varphi$
&  $(        -5.0  \pm         0.2) \, 10^{ -3}$
&  $(        5.94  \pm        0.06) \, 10^{ -5}$
\\
\end{tabular}
\label{table1}
\end{table}

\end{document}